\documentclass[twocolumn,english,aps,prl,showpacs,amsfonts,amsmath,amssymb,superscriptaddress,footinbib,floatfix,longbibliography]{revtex4-1}
\usepackage[T1]{fontenc}
\usepackage[latin9]{inputenc}
\usepackage{amstext}
\usepackage{amssymb}
\usepackage{graphicx}

\makeatletter
\usepackage{babel}
\usepackage{ulem}
\usepackage{color}

\def\kbar{{\mathchar'26\mkern-9mu k}}

\usepackage{filemod}

\makeatother

\begin{document}

\title{Return to the Origin as a Probe of Atomic Phase Coherence}

\author{Cl{\'e}ment Hainaut}

\affiliation{Universit{\'e} de Lille, CNRS, UMR 8523 -- PhLAM -- Laboratoire
de Physique des Lasers Atomes et Mol{\'e}cules, F-59000 Lille, France}

\homepage{www.phlam.univ-lille1.fr/atfr/cq}

\author{Isam Manai}

\affiliation{Universit{\'e} de Lille, CNRS, UMR 8523 -- PhLAM -- Laboratoire
de Physique des Lasers Atomes et Mol{\'e}cules, F-59000 Lille, France}

\author{Radu Chicireanu}

\affiliation{Universit{\'e} de Lille, CNRS, UMR 8523 -- PhLAM -- Laboratoire
de Physique des Lasers Atomes et Mol{\'e}cules, F-59000 Lille, France}

\author{Jean-Fran{\c c}ois Cl{\'e}ment}

\affiliation{Universit{\'e} de Lille, CNRS, UMR 8523 -- PhLAM -- Laboratoire
de Physique des Lasers Atomes et Mol{\'e}cules, F-59000 Lille, France}

\author{Samir Zemmouri}

\affiliation{Universit{\'e} de Lille, CNRS, UMR 8523 -- PhLAM -- Laboratoire
de Physique des Lasers Atomes et Mol{\'e}cules, F-59000 Lille, France}

\author{Jean Claude Garreau}

\affiliation{Universit{\'e} de Lille, CNRS, UMR 8523 -- PhLAM -- Laboratoire
de Physique des Lasers Atomes et Mol{\'e}cules, F-59000 Lille, France}

\author{Pascal Szriftgiser}

\affiliation{Universit{\'e} de Lille, CNRS, UMR 8523 -- PhLAM -- Laboratoire
de Physique des Lasers Atomes et Mol{\'e}cules, F-59000 Lille, France}

\author{Gabriel Lemari{\'e}}

\affiliation{Laboratoire de Physique Th{\'e}orique, UMR 5152, CNRS and Universit{\'e}
de Toulouse, F-31062 Toulouse, France}

\author{Nicolas Cherroret}

\affiliation{Laboratoire Kastler Brossel, UPMC-Sorbonne Universit{\'e}s, CNRS,
ENS-PSL Research University, Coll{\`e}ge de France, 4 Place Jussieu,
75005 Paris, France}

\author{Dominique Delande}

\affiliation{Laboratoire Kastler Brossel, UPMC-Sorbonne Universit{\'e}s, CNRS,
ENS-PSL Research University, Coll{\`e}ge de France, 4 Place Jussieu,
75005 Paris, France}

\begin{abstract}
We report on the observation of the coherent enhancement of the return
probability (``enhanced return to the origin'', ERO) in a periodically
kicked cold-atom gas. By submitting an atomic wave packet to a pulsed,
periodically shifted laser standing wave, we induce an oscillation
of ERO in time and explain it in terms of a periodic, reversible dephasing
in the weak-localization interference sequences responsible for ERO.
Monitoring the temporal decay of ERO, we exploit its quantum coherent
nature to quantify the decoherence rate of the atomic system. 
\end{abstract}


\maketitle
The transport of waves in disordered or chaotic systems can be strongly affected by interference effects, with striking signatures for both quantum and classical waves: 
coherent backscattering, universal conductance fluctuations~\cite{akkermans2007mesoscopic}, Anderson localization~\cite{abrahams201050} and its many-body 
counterpart~\cite{basko2006metal}.
Intuitively, one expects multiple scattering by disorder
to lead to a pseudo-random walk, and to an average diffusive behavior
at long time. For waves however, the situation is quite
different: even at moderate disorder strengths spectacular manifestations of localization can already
show up. A well known example is weak localization. In time-reversal invariant systems, two paths counterpropagating on a closed loop have the same amplitude and phase; they interfere constructively, doubling the probability to return to the starting point.

Because weak localization crucially relies on time-reversal symmetry and phase coherence, it has been exploited in many contexts to probe decoherence or magnetic field effects.
In particular, in mesoscopic electronic systems, it features a reduction of the diffusion coefficient
and constitutes an invaluable asset for probing the electronic phase
coherence~\cite{Bergman:SpinOrbitCouplingWeakLoc:PRL82,Niimi:DisorderMesoscopicWavesCoh:PRL09,Capron:DecoherenceMesoscopic:PRB13}.
In classical wave systems, weak localization is usually evidenced
by the coherent backscattering effect, which corresponds to an enhanced
probability for a wave to be reflected from a disordered medium in
the backward direction~\cite{Wolf:WeakLocCBSLight:PRL85,Albada:ObsfWeakLocLight:PRL85,Labeyrie:CBSLightColdAtoms:PRL99, PhysRevLett.97.013004}.
A third consequence of weak localization is the enhancement of the
probability that an expanding wave packet returns to its origin (``enhanced return to the origin'', ERO). This
effect manifests itself as a narrow peak visible at the center of
the density profile of the wave packet. ERO has been observed
in classical wave systems, for instance in the near-field intensity
profile of seismic waves propagating in the crust~\cite{Larose:WeakLocalizationSeismicWaves:PRL04}
or of acoustic waves in chaotic cavities~\cite{deRosny:CBSElasticWaves:PRL00,Weaver:EnhancedBackscatteringElastic:PRL00}.

\begin{figure}[!hb]
\begin{centering}
\includegraphics[width=0.9\columnwidth]{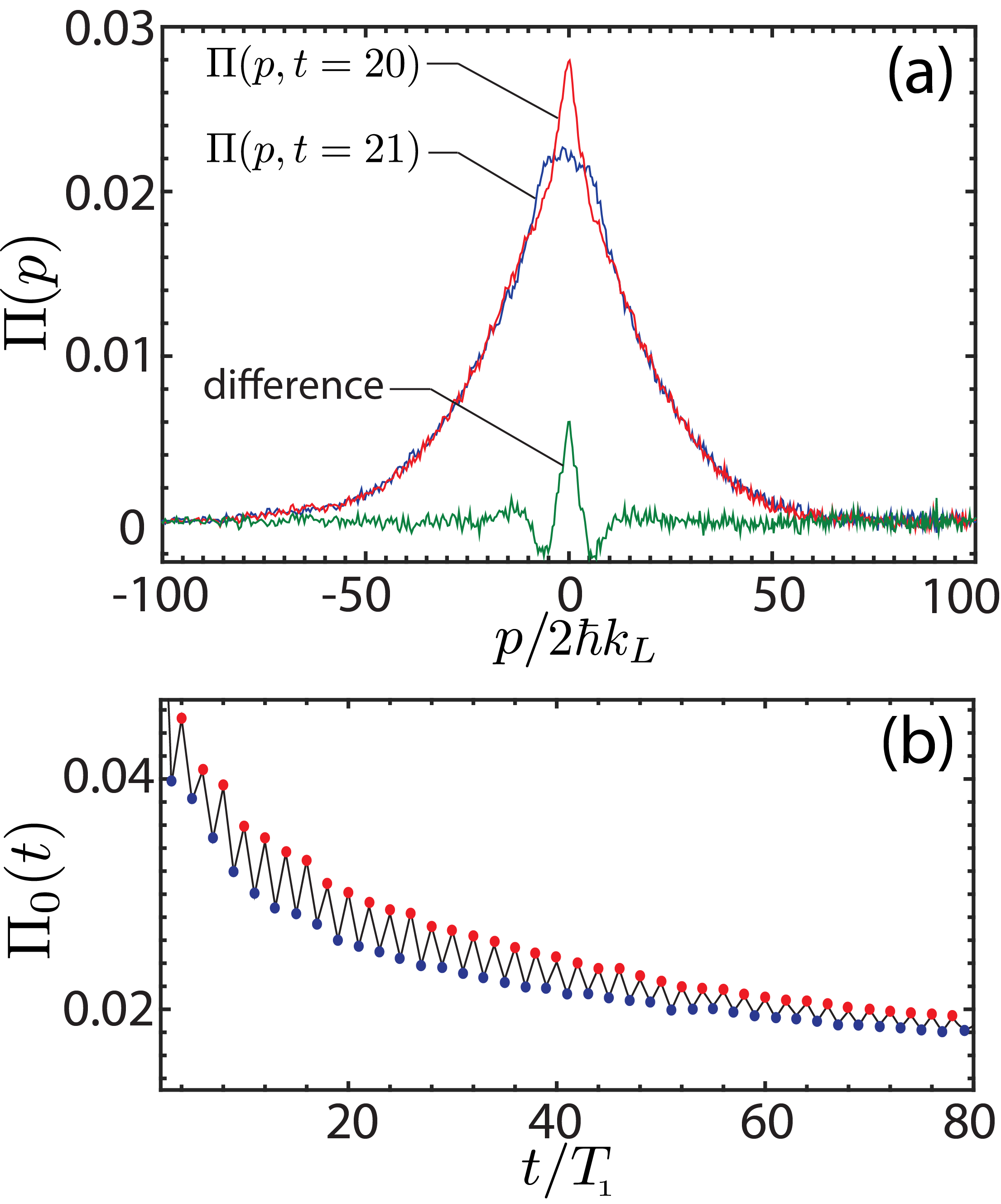} 
\par\end{centering}

\caption{\label{fig:EROobservation} 
(Color online) Experimental observation
of enhanced return to the origin. (a) Momentum distribution
$\Pi(p,t)$ at an even ($t=20$, red curve) and an odd ($t=21$, blue curve) kick. The distribution around $p\!=\!0$ at $t=20$ is enhanced
with respect to the distribution at $t=21$, as evidenced
by the green difference signal. (b) The zero-momentum population
$\Pi_{0}$ \textit{vs}. $t$ shows a clear oscillation between even kicks
(ERO, red dots) and odd kicks (blue dots). The attenuation in the contrast is due to decoherence. Parameters are $K=12$, $\kbar=1.5$ and $a=0.04$.
}
\end{figure}

Recent cold atom experiments~\cite{SanchezPalencia:DisorderQGases:NP10} offer a high level of control on crucial ingredients like statistical properties of disorder, dimensionality, interactions and coupling to the environment. This has led to
clear new observations of Anderson localization~\cite{Billy:AndersonBEC1D:N08,Roati:AubryAndreBEC1D:N08}, 
coherent backscattering~\cite{Jendrzejewski:CBSUltracoldAtoms:PRL12}, 
and recently many-body localization~\cite{schreiber2015observation}. On the other hand, the atomic quantum kicked rotor (QKR),
a model system for quantum chaos~\cite{Izrailev:LocDyn:PREP90},
has played a key role in the observation of \textit{dynamical
localization}, a suppression of the classical chaotic diffusion in
momentum space by quantum interference~\cite{Casati:LocDynFirst:LNP79,Moore:AtomOpticsRealizationQKR:PRL95},
analog to Anderson localization~\cite{Fishman:LocDynAnders:PRL82}.
By adding modulation frequencies~\cite{Shepelyansky:Bicolor:PD87,Casati:IncommFreqsQKR:PRL89}, ``quantum
simulations''~\cite{Georgescu:QuantumSimulation:RMP14} of multidimensional
Anderson models have been realized in 2D~\cite{Manai:Anderson2DKR:PRL15} and
3D~\cite{Chabe:Anderson:PRL08,Lemarie:CriticalStateAndersonTransition:PRL10,Lopez:ExperimentalTestOfUniversality:PRL12,Lopez:PhaseDiagramAndersonQKR:NJP13,Lemarie:AndersonLong:PRA09}
systems, where the metal-insulator transition has been completely characterized.

\begin{figure*}[!ht]
\begin{centering}
\includegraphics[height=5cm]{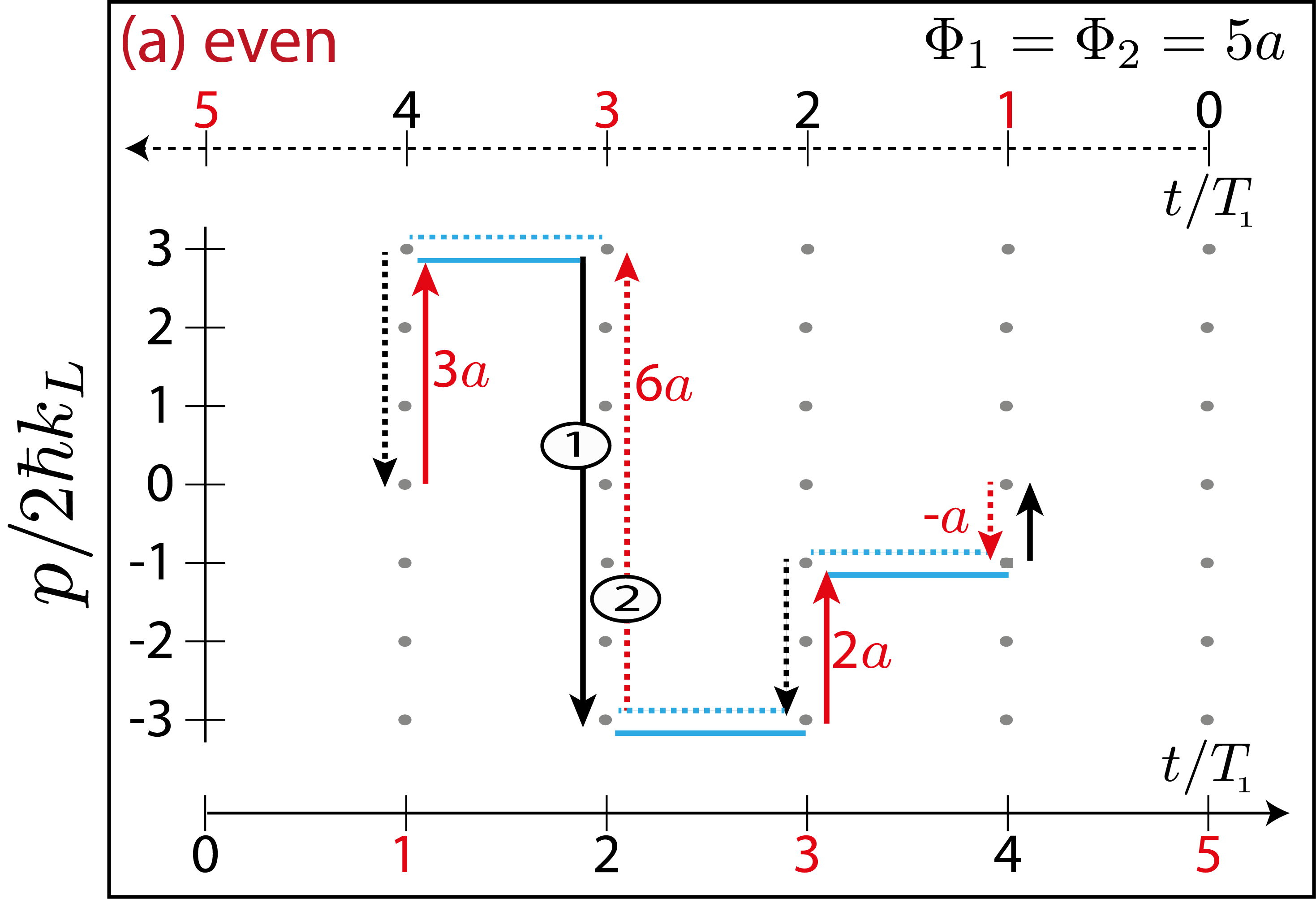}$\quad$\includegraphics[height=5cm]{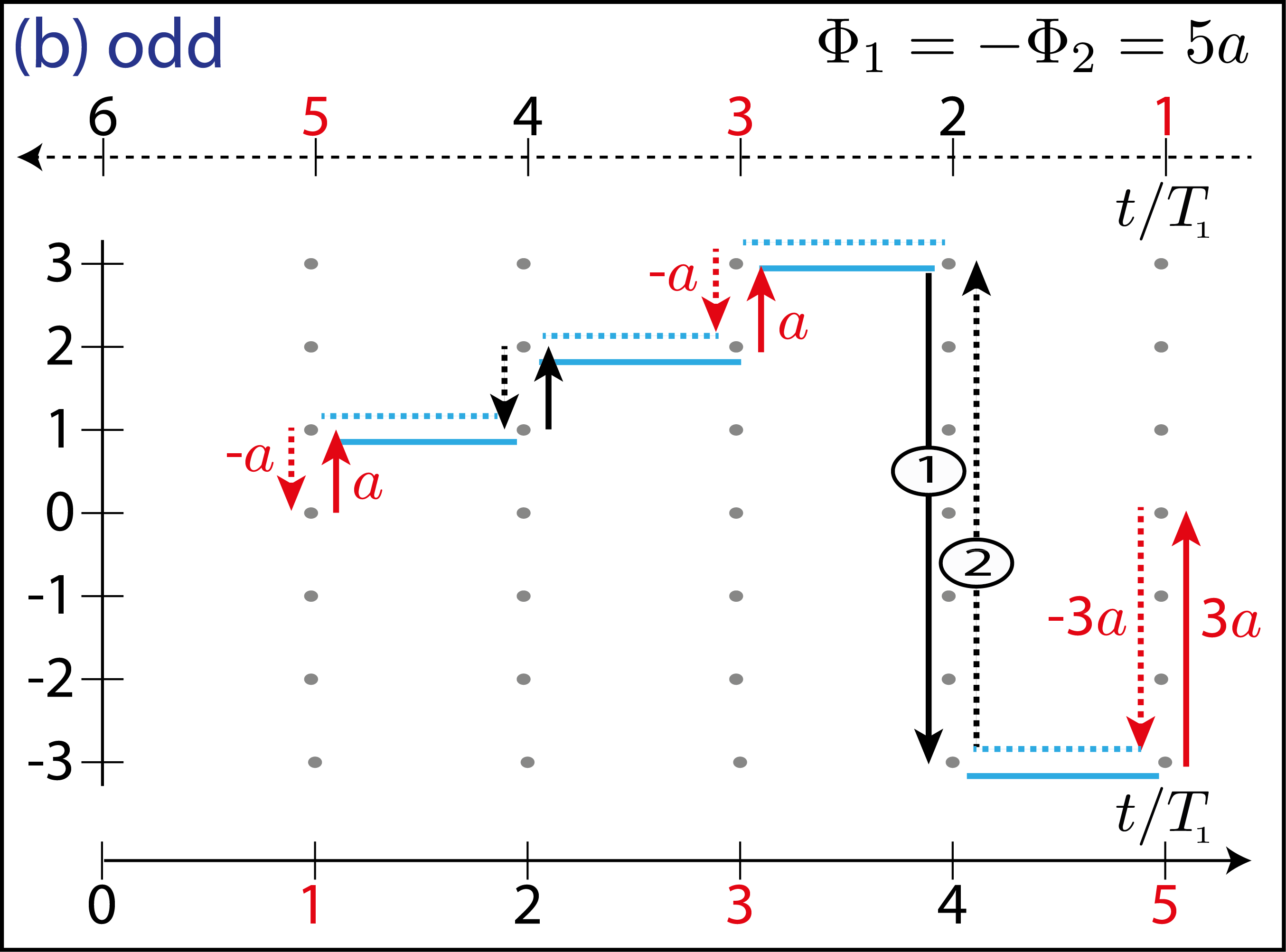}
\par\end{centering}

\caption{\label{fig:TimeReversalInv}(Color online) 
Paths in momentum space at the origin of ERO in the PSQKR.
For each path we show the amplitude (solid lines) and its time-reversed associate
(dashed lines). Kick momentum transfers are represented by vertical
arrows, red for odd kicks and black for even kicks. Arrows point in
the direction of increasing time. Horizontal lines symbolize free
propagation between kicks. (a) Paths involving four kicks: Path 1
accumulates an $a$-dependant phase $\Phi_{1}=3a+2a=5a$ associated with
the shifted SW position at first and third kicks. The time-reversed
path 2 accumulates exactly the same phase $\Phi_{2}=-a+6a=5a$, making
ERO visible. (b) Same thing for five kicks: The total phases are
respectively $a+a+3a=5a$ and $-a-a-3a=-5a$. The phase difference is
$\Delta\Phi=10a\protect\ne0$ and the ERO is suppressed.}
\end{figure*}

Experimentally, ERO is difficult to observe as it requires an initially narrow wave packet and a good spatial resolution. In this Letter, we use the \textit{full control} of the scattering events (here the kicks) that occur during the propagation of the atomic kicked rotor -- in contrast with usual disordered media where scattering events occur randomly in time -- to periodically trigger or extinguinsh the interference mechanisms at the origin of ERO. The observation of ERO is achieved through striking oscillations of the return probability. It thus constitutes a excellent probe of the "building blocks" of the interference processes leading to localization. Furthermore, by following in time the destruction of ERO, we measure the decoherence of the system, in the spirit of studies conducted in mesoscopic physics. Decoherence is nowadays recognized as a fundamental process bridging quantum physics at the microscopic scale with classical physics at the macroscopic scale ~\cite{Zurek:Decoherence:RMP03,Haroche:NobelLecturePhotonsinaBox:RMP13}.

In our experiment, a cloud of laser-cooled atoms is exposed to a pulsed,
far-detuned standing wave (SW).
A key feature is the use of a modified version of the
QKR~\cite{Tian:EhrenfestTimeDynamicalLoc:PRB05}, in which the SW
is spatially shifted every second kick by an amount $a$. We call such a system ``periodically-shifted
QKR'' (PSQKR), and it is described by the Hamiltonian
\begin{equation}
H\!=\!\frac{p^{2}}{2}+K\sum_{n}\left[\cos x\ \delta(t\!-\!2n)+\cos(x\!+\!a)\ \delta(t\!-\!2n\!+\!1)\right],\label{eq:Hpsqkr}
\end{equation}
where time is measured in units of the SW pulse period $T_{1}$, space
in units of $(2k_{L})^{-1}$ with $k_{L}=2\pi/\lambda_{L}$ the
laser wave number, and momenta in units of $2\hbar k_L$ such that $[x,p]=i\times4\hbar k_{L}^{2}T_{1}/M=i\kbar$,
defining the reduced Planck constant $\kbar$. $K$ is proportional
to the intensity and to the inverse of the detuning of the SW. Note
that, for $a=0$, Eq.~(\ref{eq:Hpsqkr}) reduces to the Hamiltonian
of the usual QKR~\cite{Casati:LocDynFirst:LNP79}.

For the kicked rotor, diffusion and localization 
take place in momentum space, hence ERO will manifest itself as a narrow peak around the initial momentum
$p \approx 0$ in the momentum density. Its observation thus requires a very good momentum resolution, both in the
measurement and in the preparation processes. The experimental ERO signal is convoluted with the width of
the initial momentum distribution, which reduces the enhancement factor well below the
expected value of 2, making its direct observation difficult. It is thus necessary to start with a momentum
distribution as narrow as possible. We load Cs atoms in a standard Magneto-Optical Trap (MOT), and cool them further by an optimized molasses phase, which cools the atoms to a temperature of $2\,\mu$K.

We then apply a pulsed optical standing wave ~\footnote{In practice the kicks have a finite duration $\tau=350$~ns. Free propagation can be neglected as long as the atom motion during this time is small compared to the scale of variation of the potential, that is $v\tau\ll\lambda_{L}/2$, where $v$ is the atom velocity and $\lambda_{L}$ the wavelength of the standing wave.}, formed by two independent laser beams ~\cite{Manai:Anderson2DKR:PRL15}. The standing wave is spatially shifted by changing the phase of one beam  with respect to the other; doing so each other kick realizes the  PSQKR described by the Hamiltonian (\ref{eq:Hpsqkr}). As this Hamiltonian is of period 2, the ERO peak is present only each second kick (see below), making its observation easier (see Fig. 1).

The atomic momentum distribution $\Pi(p,t)$ is detected by a standard time-of-flight technique at
the end of the sequence. At even kicks (to which no spatial shift
is applied) we clearly observe an enhancement of $\Pi(p)$ in the
vicinity of $p=0$ [red curve in Fig.~\ref{fig:EROobservation}(a)]
for $t=20$. In contrast, at odd kicks [$t=21$, blue curve in
Fig.~\ref{fig:EROobservation}(a)] no enhancement is visible. Fig.~\ref{fig:EROobservation}(b)
shows $\Pi_{0}(t)\equiv\Pi(p=0,t)$; one sees that this oscillatory
behavior persists up to long times $t>80$.

One can understand the origin of the oscillation of ERO in our system by considering
the PSQKR evolution operator over one time
period (corresponding to two kicks). For symmetry reasons, we choose
to consider the evolution operator $U$ from time $2n-1/2$ to $2n+3/2$.
Indeed, momentum densities do not evolve during free propagation between
kicks, so the final results do not depend on the origin of time.
This evolution operator can then be split in a
``shifted'' (odd) kick operator $U_{a}$ and a ``non-shifted'' (even) evolution
operator $U_{0}$: $U=U_{a}U_{0}$ with
\begin{eqnarray}
U_{a} & = & \exp\left(-\frac{i\hat{p}^{2}}{4\kbar}\right)\exp\left[-i\kappa\cos\left(\hat{x}\!+\!a\right)\right]\exp\left(-\frac{i\hat{p}^{2}}{4\kbar}\right)\label{eq:Ua}\\
U_{0} & = & \exp\left(-\frac{i\hat{p}^{2}}{4\kbar}\right)\exp\left[-i\kappa\cos\hat{x}\right]\exp\left(-\frac{i\hat{p}^{2}}{4\kbar}\right),\label{eq:U0}
\end{eqnarray}
where $\kappa\equiv K/\kbar$. A key point for ERO
is the existence of constructive interference between time-reversed
paths. 
In the usual QKR, this is due to the invariance of the evolution
operator over one kick -- which coincides with $U_{0}$
-- under the generalized time-reversal symmetry operator $\mathcal{T}=TP$,
product of the time-reversal anti-unitary operator $T\!:\!t\!\to\!-t$
 with the unitary parity operator $P\!:\!x\!\to\!-x$, such that 
$\mathcal{T}\!:\!t\!\to\!-t;\,x\!\to\!-x;\,p\!\to\!p$ preserves momentum. 
For the PSQKR, $\mathcal{T}=TP$ is not a symmetry
operation, because the $a$ term in $U_{a}$
is not parity-invariant. However, the product $\mathcal{T}_a=TP_{a/2}$ of the time-reversal
operator by the parity operator with respect to $a/2$, $P_{a/2}\!:\!x\!\to\!a-x$
exchanges $U_{0}$ and $U_{a}$: $\mathcal{T}_a U_{0,a}\mathcal{T}_a=U_{a,0}$ Thus, for even numbers
of kicks the symmetry is preserved: $\mathcal{T}_a(U_aU_0)^n\mathcal{T}_a=(U_aU_0)^n$, but, for odd numbers of kicks,
an orphaned $U_{0}$ or $U_{a}$
operator remains, breaking the symmetry. As a consequence, multiple
scattering paths which are images of each other by $\mathcal{T}_a$
will accumulate the same phase, leading to a constructive interference,
very much like time-reversed paths are responsible for weak localization
in usual time-reversal invariant disordered systems.

To illustrate this reasoning, let us consider an example. With periodic boundary
conditions~\footnote{In the experiment, the system is extended
along $x$ so that we cannot use periodic boundary conditions. Nevertheless,
the system is invariant by a 2$\pi$ spatial translation, so that
the Bloch theorem applies. Any initial state of the system can be
written as a linear combination of different quasimomenta $\beta\kbar$ in the
first Brillouin zone $-1/2<\beta\le1/2$. The various $\beta$ components
are uncoupled and evolve independently. $\beta$ -- like the momentum
itself -- is preserved by the $\mathcal{T}$ symmetry, so that all
$\beta$ components display the ERO phenomenon. The only change is
the replacement $n\to n+\beta$ for the phase accumulated during free
propagation, which affects similarly the pair of conjugate paths.}
along $x$, we can use the eigenbasis associated with the $\hat{p}$ operator, labeled
by an integer $n$ such that $\hat{p}|n\rangle=n\kbar|n\rangle.$
The free propagation operator in this basis is diagonal, while the kick operator
is $\exp\left[-i\kappa\cos\left(\hat{x}\!+a\!\right)\right]=\sum_{m}(-i)^{m}J_{m}(\kappa)\mathrm{e}^{ima}|n+m\rangle\langle n|$
(with $a\!=\!0$ for even kicks). For odd kicks ($a\neq0$) the side
bands generated from component $n$ get an additional phase $ma$,
where $m$ is the change in momentum. In panel (a) of Fig.~\ref{fig:TimeReversalInv}
we represent by a broken solid line a ``momentum path'' (labeled
1) involving 4 kicks, to which we
match the associated time-reversed path 2 (broken dashed line). Such
sequence of counter-propagating paths is responsible for ERO~\cite{Prigodin:MesoDynEchoQuantumDots:PRL94}.
One sees that both the direct and the time-reversed paths accumulate
the same phase (here $\Phi_{1}=\Phi_{2}=5a$). The dephasing $\Phi_{1}-\Phi_{2}$
vanishes, making ERO visible. In contrast, considering a 5 kick path and its time-reversed image, 
Fig.~\ref{fig:TimeReversalInv}b,
a residual dephasing ($\Phi_{1}-\Phi_{2}=10a$) remains, suppressing ERO.

The periodic manifestation of ERO in our
system can also be understood from the diagrammatic technique~\cite{Altland:DiagrammaticAndersonLocQKR:PRL93}.
Assuming that transport is supported by diffusion, we find
\begin{equation}
\Pi_{0}(t)\simeq\frac{1}{\sqrt{4\pi Dt}}\left[1+e^{-\Gamma t}\times\left\lbrace \begin{array}{ccc}
1 & \mbox{if}\ t\ \text{even}\\
\mathrm{e}^{-a^{2}Dt} & \mbox{if}\ t\ \text{odd}
\end{array}\right.\right],\label{ERO_eq}
\end{equation}
where $D$ is the diffusion coefficient and $\Gamma$ the decoherence
rate of the system. The second term in the square brackets is the
contribution of ERO. In agreement with the experimental observation,
at finite $a$ this contribution is strongly suppressed at odd kicks.
While Eq.~(\ref{ERO_eq})
predicts an enhancement factor of 2 between even and odd kicks for
sufficiently large $a$, the experimentally
observed factor is significantly lower, essentially due to the convolution with the initial momentum profile as discussed above.
Note also that the $t^{-1/2}$ dependance of the ERO signal is expected to be valid only in the initial diffusion stage, whereas the decay at long times is essentially dominated by exponential terms in Eq. ~(\ref{ERO_eq}) (see Figure 4).

\begin{figure}[!t]
\begin{centering}
\includegraphics[height=5cm]{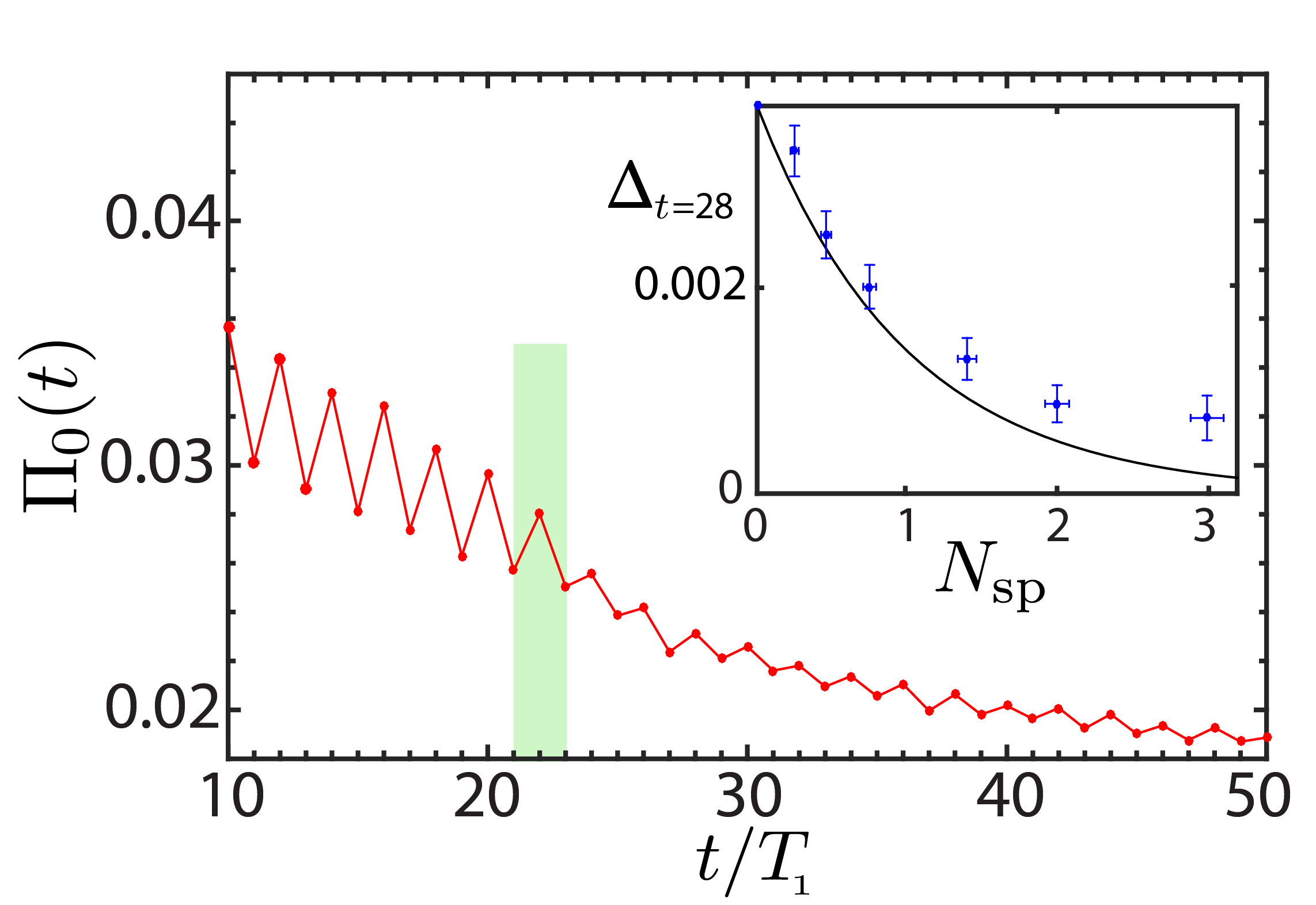} 
\par\end{centering}

\caption{\label{fig:decoherencePulse}(Color online) Zero-momentum probability
density $\Pi_{0}$ \textit{vs.} $t$
for the PSQKR.  A
decoherer beam is applied between the $21^{\mathrm{st}}$ and
$23^{\mathrm{rd}}$ pulses (green-shadowed region), quenching the
oscillations. Parameters
are $K=12$, $\kbar=1.5$ and $a=0.04$. The decoherer beam
induces an average of $N_{\text{sp}}=2$ spontaneous emission events  per
atom.
The inset shows the reduction in the difference signal $\Delta_{t}$
as a function of $N_{\text{sp}}$; the black line is the expected
exponential decay $\exp(-N_{\text{sp}})$ (it is \textit{not} a fit).}
\end{figure}

To demonstrate that the experimental ERO signal is due to quantum interference between pairs of closed loops,
we add a controlled amount of decoherence to the
system. For this purpose, we define the quantity $\Delta_{t}=\left(-1\right)^{t}\left[\Pi_{0}(t\!=\!n)-\Pi_{0}(t\!=\!n-1)\right]$,
the difference of the zero-momentum population between two successive
kicks. We shine on the atoms a resonant laser (``decoherer'') beam
at $t=21^{+}$ (i.e just after the $21^{\mathrm{st}}$ kick) of a
PSQKR sequence, thus producing spontaneous emission-induced decoherence.
The decoherer is applied during $20\mu$s (up to $t=23$) and its intensity
is adjusted to produce an average number $N_{\text{sp}}$ of spontaneous
emission events per atom. This number is independently calibrated
by shining the decoherer beam on the
MOT cloud and measuring the radiation pressure force it exerts on
the atomic sample. The effect of the decoherer beam on the ERO
signal is shown in Fig.~\ref{fig:decoherencePulse}: the oscillating behavior of $\Pi_{0}$ is rapidly quenched 
after kick 21, which proves the coherent nature of the observed
ERO. The inset of Fig.~\ref{fig:decoherencePulse} shows the decrease
of $\Delta_{t=28}$ \textit{vs}. $N_{\text{sp}}$, displaying an exponential behavior $\exp(-N_{\text{sp}})$.
Indeed, ERO still exists \textit{after} the decoherer
pulse, due to atoms which have not scattered any
resonant photon, and, as this is a Poissonian process, the probability
of scattering zero photon is $\exp(-N_{\text{sp}})$.
The remaining small $\Delta_{t=28}$ at large $N_{\text{sp}}$ is probably due to the incomplete quenching
of phase coherence by spontaneous emission.

\begin{figure}[!t]
\begin{centering}
\includegraphics[height=5cm]{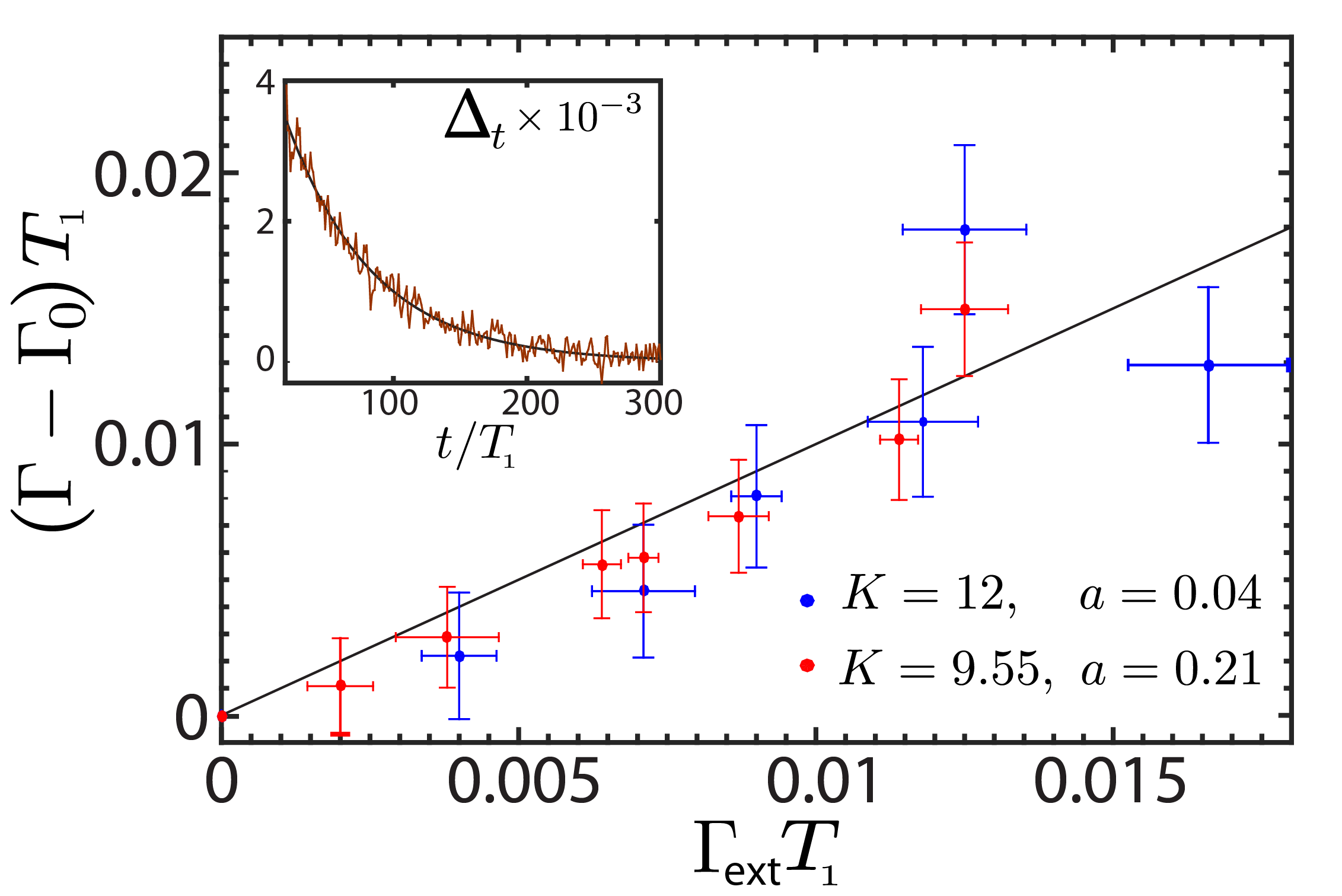} 
\par\end{centering}

\caption{\label{fig:decoherenceContinuous}(Color online) Probing decoherence
with ERO. Inset: the decay of the difference signal $\Delta_{t}$
\textit{vs.} $t$ is fitted by an exponential (black line) from which the
decoherence rate $\Gamma$ is extracted.
In the absence of any externally applied decoherence, this gives the
-- $K$ and $a$ dependent -- "stray" decoherence rate $\Gamma_0.$ This
procedure is repeated in the presence of the decoherer beam for several
values of the additional imposed decoherence rate
$\Gamma_{\mathrm{ext}}$. The fact that the excess rate
$\Gamma-\Gamma_0$ measured using the decay of the ERO signal agrees
perfectly with the externally added rate $\Gamma_{\mathrm{ext}}$ shows
that ERO is a faithful measure of decoherence.}
\end{figure}

The ERO signal can also be used to \textit{measure} the amount of decoherence
present in the system.
We observe an exponential decay of $\Delta_{t}$ \textit{vs.} $t$, shown
in the inset of Fig.~\ref{fig:decoherenceContinuous}, from which
one can determine the decoherence rate $\Gamma_0$:
$\Gamma_{0}=0.024$ for $K=12$ and $\Gamma_{0}=0.014$ for $K=9$.
Which physical mechanisms induce this decoherence is presently unknown~\footnote{See
Ref.~\cite{Lemarie:AndersonLong:PRA09} for a more complete discussion
of the possible decoherence sources in our setup.}.
We can nevertheless test the reliability of the method by applying the decoherer beam
during the \textit{whole} experimental sequence, thus introducing a controlled
amount of spontaneous emission. The beam intensity, calibrated \textit{in situ}
by measuring the radiation pressure on the atomic cloud as described above, is chosen
to produce a controlled decoherence rate $\Gamma_{\mathrm{ext}}$. 
From the decay of $\Delta_{t}$ \textit{vs.} $t$, 
we determine the total decoherence rate $\Gamma.$  We expect the latter to be given by
$\Gamma=\Gamma_{\mathrm{ext}}+\Gamma_{0.}$ The straight line of slope 1
in Fig.~\ref{fig:decoherenceContinuous} (\textit{not a fit}) proves
that it is indeed the case, so that we have a reliable measurement of decoherence rates,
very much like magnetoconductance is used in solid state physics to measure the electronic phase coherence length~\cite{Bergman:SpinOrbitCouplingWeakLoc:PRL82,Niimi:DisorderMesoscopicWavesCoh:PRL09,Capron:DecoherenceMesoscopic:PRB13}. 

In conclusion, we have experimentally observed the phenomenon of enhanced return to the origin with atomic matter waves,
a clear signature of weak localization in time-reversal invariant systems.
By controlling the phase of the scattering events induced by the standing wave kicks, we have induced
a time-periodic oscillation of ERO, allowing for a clear observation of its contrast. A crucial ingredient
is the ability to control precisely the even/odd number of scattering events, a unique advantage of the kicked rotor,
in contrast with ordinary disordered systems where only the average number of scattering events is under control.
Finally, by introducing
a controlled amount of decoherence, we have verified
its quantum nature and used it to access the
decoherence rate in the system.
This work opens promising perspectives
in the use of coherent phenomena to probe sources of decoherence in
atomic systems, as well as other sources of dephasing such as interactions~\cite{Hartung:CBSBEC2D:PRL08}. 
Phase control of scattering events may also constitute an alternative approach to 
artificial gauge fields~\cite{Dalibard:ArtificialGaugePotentials:RMP11}
to induce effective magnetic field effects in cold atom systems.

\begin{acknowledgments}
The authors are grateful to C. Tian for fruitful discussions.
This work is supported by Agence Nationale de la Recherche (Grants
LAKRIDI No. ANR-11-BS04-0003-02 and K-BEC No. ANR-13-BS04-0001-01),
the Labex CEMPI (Grant No. ANR-11-LABX-0007-01), and ``Fonds Europ{\'e}en
de D{\'e}veloppement Economique R{\'e}gional'' through the ``CPER Photonics for Society (P4S)''. This work was granted access to the HPC resources of TGCC under the allocation 2016-057083 made by GENCI (Grand Equipement National de Calcul Intensif)
\end{acknowledgments}


%

\end{document}